\documentclass{ws-procs975x65}

\begin{document}

\title{VARIABLE TENSION BRANE-WORLDS}

\markboth{Gergely}{Variable tension brane-worlds} 
%\wstoc{Variable tension brane-worlds}{Gergely}

\author{L\'{A}SZL\'{O} \'{A}. GERGELY}

\address{Departments of Theoretical and Experimental Physics, University of Szeged, Hungary\\
\email{gergely@physx.u-szeged.hu}}

\begin{abstract}
I present recent work on codimension one brane-world models containing a 3+1
dimensional curved brane with time-dependent brane tension.
\end{abstract}


\bodymatter\bigskip 

According to classical (weak-field) tests of general relativity,
gravitational dynamics is well described by the Einstein equation. Testing
the stronger gravity sector by direct detection of gravitational waves is an
ongoing effort. As general relativity predicts unavoidable singularities
both in cosmological and astrophysical setups, there is hope that suitable
modifications of the theory would led to singularity avoidance. Both loop
quantum gravity\cite{loop} and quantum geometrodynamics\cite{Kiefer} have
been successful in such attempts in particular cases. Furthermore, the
confrontation of general relativity with observations both at galaxy and
galaxy cluster scale, and at cosmological scale led to the necessity of
introducing the strong energy condition violating dark energy and overall
invisible dark matter concepts. The hierarchy problem obstructs attempts to
unify all basic interactions.

For all these reasons an overall effort was made to study gravity theories,
which are different from general relativity, but they reduce to it in the
appropriate limit. In this context we mention MOND\cite{Milgrom} and it's
relativistic generalization\cite{Bekenstein}, TeVeS theories\cite{TeVeS}, $%
f(R)$ gravity theories\cite{fR} and various versions of brane-world
theories. Among them we find codimension two (or higher) branes\cite{6brane}
on the tip of a conical singularity; codimension one models containing two
branes, with only one of them physical\cite{RS1}; induced gravity models,
with a piece of Einstein-Hilbert action present both in 5 dimensions (5d)
and in 4d\cite{Induced}. The simplest such theory contains one 3+1
hypersurface (the brane) embedded \ in a 4+1 dimensional non-compactified
space-time governed by Einstein's gravity\cite{RS2}. The dynamics induced on
such a brane is known both in a covariant form\cite{BraneCovariant} and as
part of a 3+1+1 decomposition\cite{Brane3+1+1}.

One key element in brane-world theories is the brane tension $\lambda $,
related to the deficit angle of the conical singularity in 6d theories and
to the discontinuity of the extrinsic curvature in the 5d case (through the
Lanczos equation). There are various established constraints on the brane
tension\cite{GK}, all pointing towards extremely high values as compared to
nowadays observable energy densities.

Recently a yet new scenario was proposed for modifying general relativity,
by allowing for a varying $\lambda $ in the context of the codimension one
single brane scenarios\cite{VarBraneTensionPRD}, inspired by the
temperature-dependence of the fluid membrane tension\cite{Eotvos}. In such a
theory the tensorial degrees of freedom of gravity on the brane are still
governed by an effective Einstein equation: 
\begin{equation}
G_{ab}=-\Lambda g_{ab}+\kappa ^{2}T_{ab}+\widetilde{\kappa }^{4}S_{ab}-%
\overline{\mathcal{E}}_{ab}+\overline{L}_{ab}^{TF}+\overline{\mathcal{P}}%
_{ab}\ .  \label{modEgen}
\end{equation}%
[Here the source terms $S_{ab}$, $\overline{\mathcal{E}}_{ab}$, $\overline{L}%
_{ab}^{TF}$, and $\overline{\mathcal{P}}_{ab}$ are a quadratic expression in
the energy-momentum tensor $T_{ab}$, the electric part of the 5d Weyl
tensor, an asymmetric embedding generated source term, and the pull-back of
any nonstandard model 5d field with $\widetilde{T}_{ab}$, respectively; $%
\Lambda $ and $g_{ab}$ are the cosmological constant and metric on the brane
(defined by its normal $n^{a}$); finally $\kappa ^{2}$, $\widetilde{\kappa }%
^{2}$ represent the gravitational coupling constants on the brane and in
5d.] The brane tension enters both $\Lambda $ and $\kappa ^{2}$, therefore
these quantities also vary. The vectorial and scalar gravitational degrees
of freedom are governed by the Codazzi and twice contracted Gauss equation%
\cite{VarBraneTensionPRD}. From among them, solely the energy-balance
equation obtained as the difference ($\Delta $) of the Codazzi equations
taken on the two sides of the brane, acquires a new term: 
\begin{equation}
\nabla _{c}T_{a}^{c}=\nabla _{a}\lambda -\Delta (g_{a}^{c}{}n^{d}{}%
\widetilde{T}_{cd})\ .  \label{en_balance}
\end{equation}

The cosmological evolution of a perfect fluid was also investigated\cite%
{EotvosBrane}. Due to the imposed cosmological symmetries, $\lambda $, $%
\kappa ^{2}$ and $\Lambda $ all become scale-factor (or cosmological time)
dependent. Borrowing the particular temperature dependence of the E\"{o}tv%
\"{o}s law\cite{Eotvos} and imposing the continuity equation (such that the
temperature dependence of the brane tension is balanced by the energy
interchange between the brane and the 5d space-time), a series of
consequences stem out:

i) The brane is formed in a very hot early universe when $\lambda =0$.

ii) Initially both $\lambda $ and $\kappa ^{2}$ were small, enhancing
brane-world effects.

iii) For small values of the scale factor $\Lambda _{early}<0$ holds,
contributing to mutual attraction; for large scale factor $\Lambda _{late}>0$
generates dark energy type repulsion.

The confrontation of this toy model with cosmological observations rendered
the spectacular part of the evolutions to the very early universe preceding
BBN, after which $\lambda $, $\kappa ^{2}$ and $\Lambda $ asymptote to
constant values. Still, further exploration of the new degree of freedom
represented by the varying brane tension by allowing for energy interchange
and other type of evolutions could led to interesting new physics. In
particular, brane-worlds have turned useful in replacing dark matter with
geometric effects, both at the galactic\cite{BraneRotCurves} and galaxy
cluster\cite{BraneGalClusters} scales. Therefore the variable brane tension
generalization of spherically symmetric brane-world solutions\cite%
{BraneStars}; also of the gravitational collapse on the brane\cite%
{BraneCollapse} may turn challenging.

{\small Participation of L\'{A}G at the MG12 Conference was made possible by
the London South Bank University Research Opportunities Fund. Additional
support from OTKA grant 69036 and NKTH Pol\'{a}nyi Program is acknowledged.}

\end{document}